\documentclass[12pt]{iopart}

\usepackage{graphicx}
\usepackage[sort&compress,square]{natbib}

\newcommand{\hii}{H~{\sc ii}}

\begin{document}

\title{The physics of volume rendering}

\author{Thomas Peters}

\address{Institut f\"{u}r Computergest\"{u}tzte Wissenschaften, Universit\"{a}t Z\"{u}rich,
Winterthurerstrasse 190, CH-8057 Z\"{u}rich, Switzerland}
\eads{tpeters@physik.uzh.ch}

\begin{abstract}
Radiation transfer is an important topic in several physical disciplines, probably most prominently in astrophysics.
Computer scientists use radiation transfer, among other things, for the visualisation of complex data sets with direct volume rendering.
In this note, I point out the connection between physical radiation transfer and volume rendering, and I describe an implementation
of direct volume rendering in the astrophysical radiation transfer code RADMC-3D. I show examples for the use of this module
on analytical models and simulation data.
\end{abstract}


\section{Introduction}

Radiation transfer plays an important role in several areas of the physical sciences such as atmospheric science, medical physics and astrophysics. Less
well known, at least in the physics community, are the various applications of radiation transfer theory in computer graphics. Computer scientists
apply physically-inspired radiation transfer not only to create photo-realistic images and animations \citep{phahum,foley3}, but also use it in a non-obvious way
for visualisation \citep{preim,sabella,drebin}. Volume rendering is a visualisation technique in which radiation transfer is quintessential and, besides being of immense
practical importance, can be very easily understood from physical principles.

In this short note, I introduce the method of direct volume rendering to an audience with a physical background. I provide a dictionary of technical
terms used by computer scientists and derive the volume rendering method from the integral form of the well-known radiation transfer equation. I then describe
an implementation of direct volume rendering in the freely available astrophysical radiative transfer code RADMC-3D\footnote{http://www.ita.uni-heidelberg.de/$\sim$dullemond/software/radmc-3d/}.
I conclude with some exemplary applications of this code to analytical models and simulations in astrophysics.

\section{The Radiative Transfer Equation and Transfer Functions}

The starting point for our considerations is the (one-dimensional) \emph{formal radiative transfer equation in integral form} \citep{rybickietal79,mihalasetal84,shu1,padmana1,peraiah02,salby12,nelson}
\begin{equation}
\label{eq:frt}
I_\nu(s_1) = I_\nu(s_0) \exp(-\tau_\nu(s_0, s_1)) + \int_{s_0}^{s_1} j_\nu(s') \exp(-\tau_\nu(s',s_1))\,\mathrm{d}s'
\end{equation}
Here, $I_\nu(s)$ is the \emph{intensity} of radiation at location $s$, $j_\nu(s)$ the \emph{emissivity},
\begin{equation}
\tau_\nu(s_0,s_1) = \int_{s_0}^{s_1} \alpha_\nu(s')\,\mathrm{d}s'
\end{equation}
the \emph{optical depth} between $s_0$ and $s_1$, and $\alpha_\nu(s)$ is the \emph{opacity} (or \emph{extinction coefficient}). All these quantities depend on the \emph{frequency} $\nu$ as well.
The radiation transfer equation~\eref{eq:frt} is easy to interpret. The radiation that arrives at $s_1$ has two contributions. The first term represents the incoming radiation at $s_0$,
which gets diluted by absorption along the ray from $s_0$ to $s_1$. The second contribution describes the emission of radiation by the material between $s_0$ and $s_1$,
which gets diluted as well until it arrives at $s_1$. Scattering processes are excluded from the formal transfer equation considered here, but can be described by the well-known \emph{rendering
equation}\footnote{In computer graphics, the term 'rendering' has two distinct meanings. The general meaning is the generation of an image from objects (\emph{models})
arranged in a \emph{scene}. More specifically, 3D rendering means the calculation of (often photorealistic) images from wire frame models. } in more
advanced treatments \citep{immel,kajiya,kajiya2}. To indicate the omission of scattered radiation, the simple rendering technique employed here is also called \emph{emission-absorption model}. In the examples
considered below, $I_\nu(s_0)$ is set to zero at the domain boundary $s_0$, so that \eref{eq:frt} only contains the integral term, which physically means that objects are visualised through their own emission
and not through their shadowing of some background illumination.

In physics, $j_\nu$ and $\alpha_\nu$ can describe, e.g., the thermal emission and absorption of dust or of molecular lines. In computer science, however,
$j_\nu$ and $\alpha_\nu$ are designed to visualise volumetric data sets. The basic idea is to send a ray from a camera through the computational domain for each 
pixel of the image. This procedure is called \emph{ray tracing}. Along each ray \eref{eq:frt} is then solved. Of course, now $j_\nu$ and $\alpha_\nu$ do not represent
any physical processes. Instead, they are defined using \emph{transfer functions}, and the assignment of a transfer function to a particular grid cell is called
\emph{classification}. For example, we might want to visualise a three-dimensional density field $\rho$.
The transfer function $T$ then specifies how much radiation a grid cell (or \emph{voxel}) in a certain density interval will emit or absorb. To display a density isosurface around
$\rho_0$, we will choose $T$ such that it is non-zero only in a small interval around $\rho_0$. Typical examples for isosurface transfer functions are
rectangle functions
\begin{equation}
T(\rho) = \cases{ c  & for $|\rho - \rho_0| \leq \varepsilon$\\
0 &  otherwise}
\end{equation}
or narrow Gaussians
\begin{equation}
\label{eq:Gauss}
T(\rho) = c \exp\left[-\left(\frac{\rho - \rho_0}{\sigma}\right)^2\right] .
\end{equation}
One then defines the emissivity $j(s) = T_j(\rho(s))$ and the extinction coefficient $\alpha(s) = T_\alpha(\rho(s))$, with the transfer functions for emission and absorption, $T_j$ and $T_\alpha$, respectively.
The choice of a proper transfer function for a given data set
is the central problem of \emph{direct volume rendering}\footnote{The attribute 'direct' refers to the fact that one works with the data itself. Alternatively,
e.g., one can map the volumetric data to a set of polygons and apply the marching cubes algorithm \citep{lorencline} to visualise an isosurface.} and in general very difficult.
In fact, this is a very active research area with important applications in, e.g., medical visualisation \citep{pfister}.

\section{From Intensities to Colours}

There are different ways to convert the results of \eref{eq:frt} to a colour image. The easiest and least interesting one is to compute \eref{eq:frt} only for a single
frequency. A \emph{colour table} can then be used to map each intensity value to a colour. The obvious problem is that integration of the ray over diverse objects in the scene, represented by distinct transfer
functions, can lead to the same final intensity value. This method is therefore fundamentally unable to assign a certain colour to a particular object. However, this ability is
of course exactly what makes volume rendering so interesting for applications in medical imaging and visualisation.

Another and more difficult option is to assign a colour spectrum to each transfer function. For example, if we describe a colour by its RGB (red, green, blue) values, then
we can define transfer functions $T_\mathrm{R}$, $T_\mathrm{G}$, $T_\mathrm{B}$ for each object and solve \eref{eq:frt} for the three frequencies $\nu = \mathrm{R}, \mathrm{G}, \mathrm{B}$.
In general, this means that the three emissivities $j_\mathrm{R}$, $j_\mathrm{G}$, $j_\mathrm{B}$ and the three opacities $\alpha_\mathrm{R}$, $\alpha_\mathrm{G}$, $\alpha_\mathrm{B}$ must be specified.
However, one generally assumes that all opacities are equal (\emph{grey opacities}), so that only one opacity $\alpha$ is used. The final intensities $I_\mathrm{R}$, $I_\mathrm{G}$, $I_\mathrm{B}$
can then directly be interpreted as RGB colour values.

As remarked above, finding appropriate transfer functions can be highly non-trivial. We illustrate the problem with the simple example of a density isosurface, modeled with
the Gaussian transfer function~\eref{eq:Gauss}. This isosurface has a certain width around $\rho_0$ in physical space (spatial distance between grid cells) and in density space
(the parameter $\sigma$ from \eref{eq:Gauss}). The parameters $\sigma$ and $c$ must be chosen such that integration of the intensities $j_\mathrm{R,G,B}$ yields
the desired colours $I_\mathrm{R,G,B}$. However, the spatial width of the isosurface does not have to be the same everywhere. For example, if the isosurface is wider for one ray than for
another ray, its colour can be darker or brighter, depending on whether the ray picks up more emission or less. Therefore, an isosurface of varying thickness cannot be
visualised with a constant transfer function if all rays should yield the same colour. Methods to compute transfer functions locally based on the variation of the data
have been developed to overcome this problem \citep{levoy,kindlmann}. We here only describe the most basic algorithm, corresponding to the current implementation in RADMC-3D.

\section{Raytracing through Volumetric Data}

Once the transfer functions for emission and absorption are defined, \eref{eq:frt} must be numerically integrated on the grid \citep{hege}. Since \eref{eq:frt} holds for all points $s_0$ and $s_1$ along the ray,
we can decompose the ray into intervals (not necessarily of equal length) $[s_{k-1},s_k]$ and arrive at 
\begin{equation}
\label{dfrt}
I_\nu(s_k) = I_\nu(s_{k-1}) \exp(-\tau_\nu(s_{k-1}, s_k)) + \int_{s_{k-1}}^{s_k} j_\nu(s') \exp(-\tau_\nu(s',s_{k}))\,\mathrm{d}s'
\end{equation}
for neighbouring grid cells at position $s_{k-1}$ and $s_k$. With the abbreviations
\begin{equation}
\theta_k = \exp(-\tau_\nu(s_{k-1}, s_k))
\end{equation}
and
\begin{equation}
b_k = \int_{s_{k-1}}^{s_k} j_\nu(s') \exp(-\tau_\nu(s',s_{k}))\,\mathrm{d}s'
\end{equation}
we can write \eref{dfrt} more conveniently as
\begin{equation}
\label{sdfrt}
I_\nu(s_k) = I_\nu(s_{k-1}) \,\theta_k + b_k .
\end{equation}
The quantity $\theta_k$ is called \emph{transparency} of the medium between the grid cells $s_{k-1}$ and $s_k$. Note that $\theta_k$ can only attain values between $0$ and $1$, corresponding to totally
opaque (large $\tau$) and fully transparent (vanishing $\tau$) material, respectively.

Applying \eref{sdfrt} recursively from the last grid cell $s_n$ back to the first one $s_0$, we get
\begin{eqnarray}
I_\nu(s_n) &= I_\nu(s_{n-1}) \,\theta_n + b_n\\
&= \Big[I_\nu(s_{n-2}) \,\theta_{n-1} + b_{n-1} \Big] \,\theta_n + b_n\\
&= I_\nu(s_0) \,\theta_1 \cdots \theta_{n-1} \,\theta_n + b_1 \,\theta_2 \cdots \theta_{n-1} \,\theta_n + \cdots + b_{n-1} \theta_n + b_n .
\end{eqnarray}
Introducing $b_0 = I_\nu(s_0)$, which we set to zero, this expression can be written shorter as
\begin{equation}
\label{sum}
I_\nu(s_n) = \sum_{k = 0}^n b_k \prod_{j = k + 1}^n \theta_j .
\end{equation}
In this form, the sum \eref{sum} is evaluated starting from the far side of the domain, and then the summation proceeds towards the camera (\emph{back-to-front}).
However, with the help of an auxiliary variable to store the accumulated transparency, \eref{sum} can be integrated in the other direction (\emph{front-to-back}) \citep{hege}.
This has the advantage to allow for \emph{early ray termination} when the transparency gets very small, so that more distant voxels cannot be seen anymore.

To compute \eref{sum}, the quantities $\theta_k$ and $b_k$ must be calculated. This requires the numerical integration of the corresponding integrals. Usually, interpolation
is used to carry out the quadrature. RADMC-3D supports both first and second order integration methods. The first order scheme assumes that $\alpha_\nu$ and
$j_\nu$ are constant in each cell, so that the integrals can be easily computed analytically.
To apply the second order method, $\alpha_\nu$ and $j_\nu$ are first mapped from the cell centres to the cell
corners. These vertex values are computed by averaging over all cells that touch the vertex. When a ray traverses a cell, the entry and exit values of $\alpha_\nu$ and $j_\nu$
at the position where the ray pierces through the cell faces are determined using bilinear interpolation from the four vertex values of each face. The second order
method than integrates the formal radiative transfer equation assuming that $\alpha_\nu$ and $j_\nu$ vary linearly between the entry and exit values, and in this case
an analytical solution is also possible \citep{olson87}.

RADMC-3D interpolates $\alpha_\nu$ and $j_\nu$ during the integration of the radiative transfer equation over a single grid cell. This means that the transfer functions
are only evaluated once for each grid cell. Computer scientists call this method \emph{pre-classification}. A better method, named \emph{post-classification}, instead interpolates
the grid variables and applies the transfer functions to the interpolated data. The latter method is known to yield better results and to reduce aliasing artifacts \citep{kaehler}.
However, since RADMC-3D is an astrophysical radiative transfer code and many grid variables can contribute to the emission and extinction coefficients, a general interpolation
of all these variables is prohibitively memory consuming. In practise, aliasing artifacts are visible but not severe with second order integration, but it can be difficult to visualise isosurfaces when the
grid variable varies a lot between neighbouring cells.

\section{Experiments with Analytical Models}

To illustrate the method, we apply it to a very simple analytical model of a prestellar core. We use a cubical volume with a box length of $0.25\,$pc, which we resolve with
$100$ grid cells in each direction. We let the number density vary as $n = 28$\,cm$^{-3} (r / $pc$)^{-2}$, where $r$ is the radial distance from the centre of the box.
We will visualise two isosurfaces. Since the number density follows a power law, we work with the logarithm $\log_{10} n$ instead of the actual density $n$
to get smooth results. We define transfer functions of the form
\begin{equation}
T(n) = \frac{h}{w} \exp\left[-0.5 \left(\frac{\log_{10} n - \log_{10} n_0}{\sigma}\right)^2\right] .
\end{equation}
The normalisation constant $w = 2.5 \times 10^{16}\,$cm is of the order of the size of a grid cell and chosen such that the integration over the Gaussian
results in the desired colour. Experimentation is required to find suitable values for $w$. 
The amplitude $h$ is different for the four transfer functions $j_\mathrm{R}$, $j_\mathrm{G}$, $j_\mathrm{B}$ and $\alpha$.
For the first isosurface, we set $\log_{10} n_0 = 3.5$, $\sigma = 0.05$ and $(h_\mathrm{R}, h_\mathrm{G}, h_\mathrm{B}, h_\alpha) = (0.4,0.4,0.8,0.001)$ (purple
with almost no absorption), for the second isosurface we pick 
$\log_{10} n_0 = 4.0$, $\sigma = 0.1$ and $(h_\mathrm{R}, h_\mathrm{G}, h_\mathrm{B}, h_\alpha) = (0.89,0.47,0.06,0.001)$ (ochre).

\begin{figure}
\includegraphics[width=145pt]{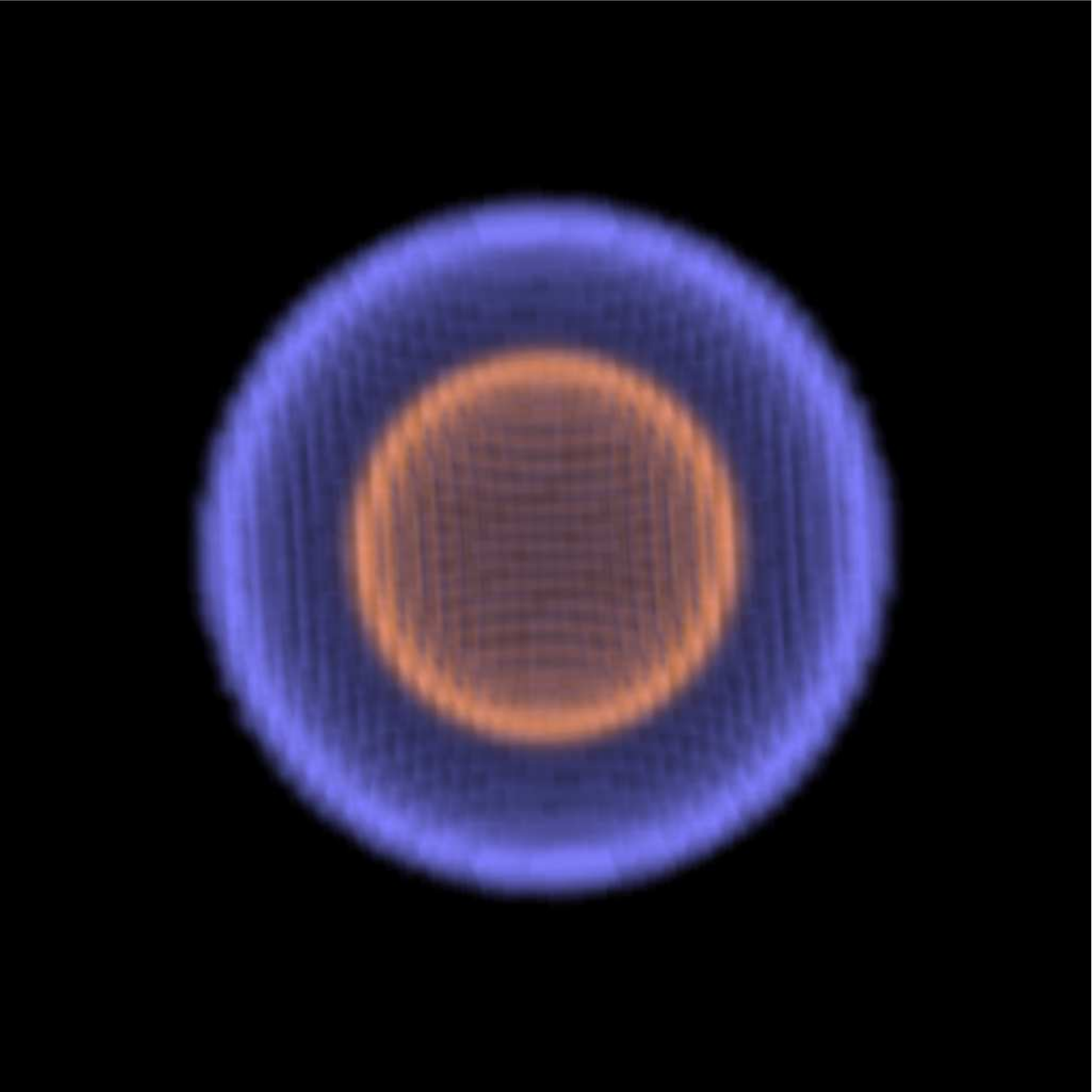}
\includegraphics[width=145pt]{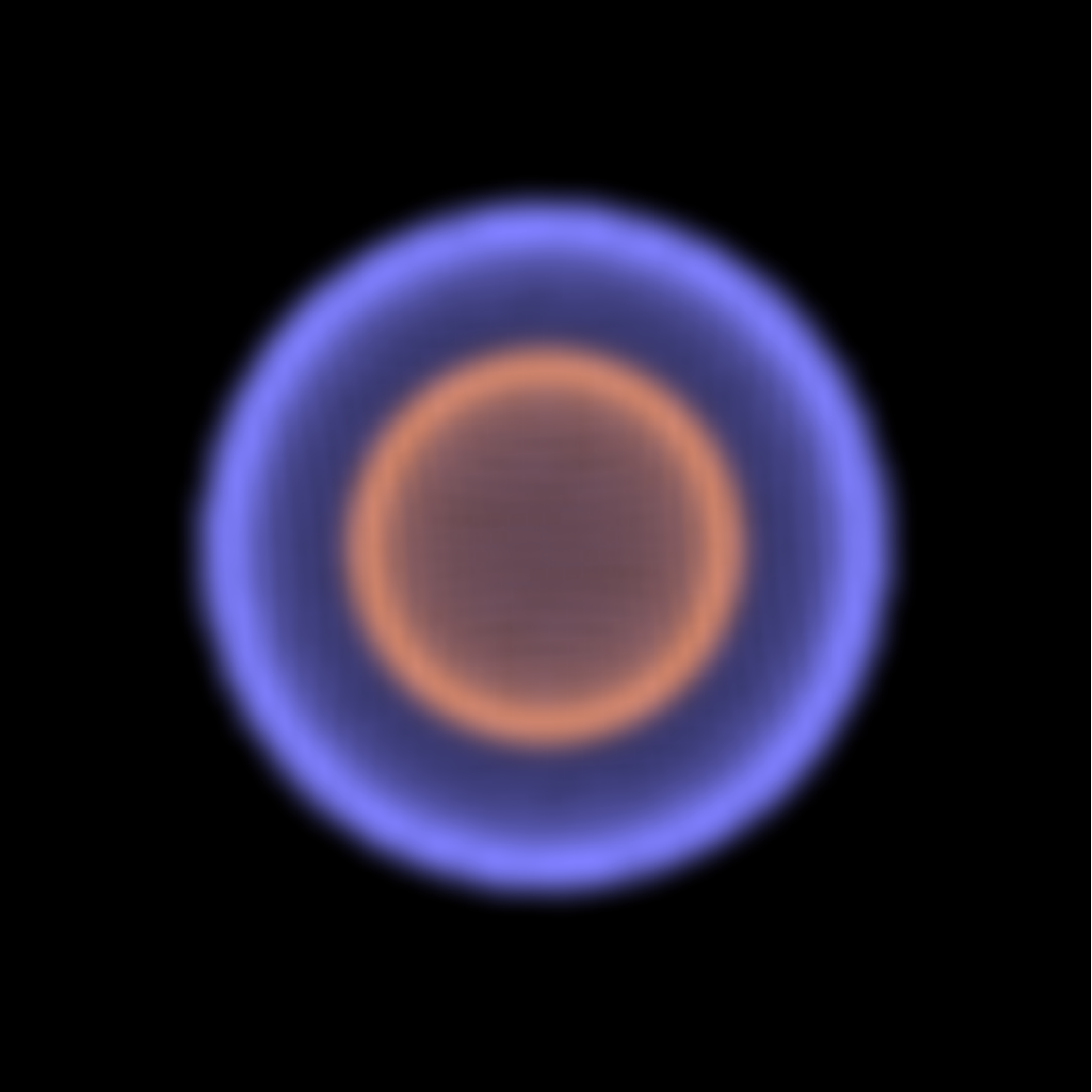}
\includegraphics[width=145pt]{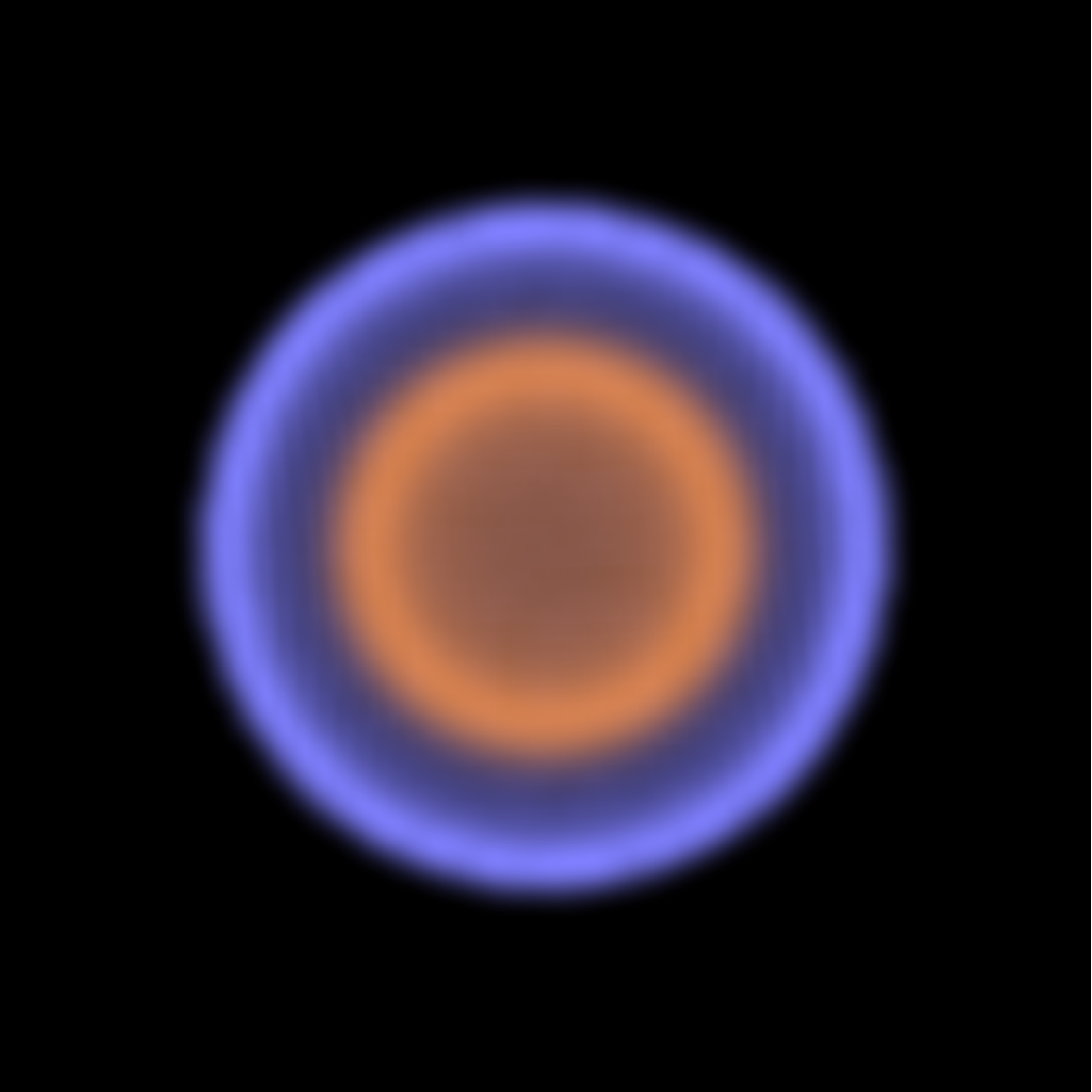}
\caption{Isosurfaces for an analytical model. The panels show, from left to right, the images generated using first order integration, second order integration,
and second order integration with an enhanced opacity for the inner isosurface.}
\label{fig:anadata}
\end{figure}

The first panel of Fig.~\ref{fig:anadata} shows the resulting image using first order integration. We see the two spherical isosurfaces, which correspond
to the different densities, with their respective colours. Since the rays pick up more contributions from the transfer functions when they hit
the isosurface tangentially, the boundaries of the spheres are much brighter than their interiors. This is analogous to, e.g., giant \hii\, region bubbles in 
the interstellar medium, where the boundaries are visible by exactly the same mechanism. We also see pronounced grid artifacts in the image. The middle
panel of Fig.~\ref{fig:anadata} displays the result of the second order integration. Here, these artifacts are much reduced. The third panel
illustrates the effect of setting $\alpha = 0.1$ for the inner isosurface, which makes this isosurface more opaque.

\section{Visualisation of Simulation Data}

\begin{figure}
\includegraphics[width=220pt]{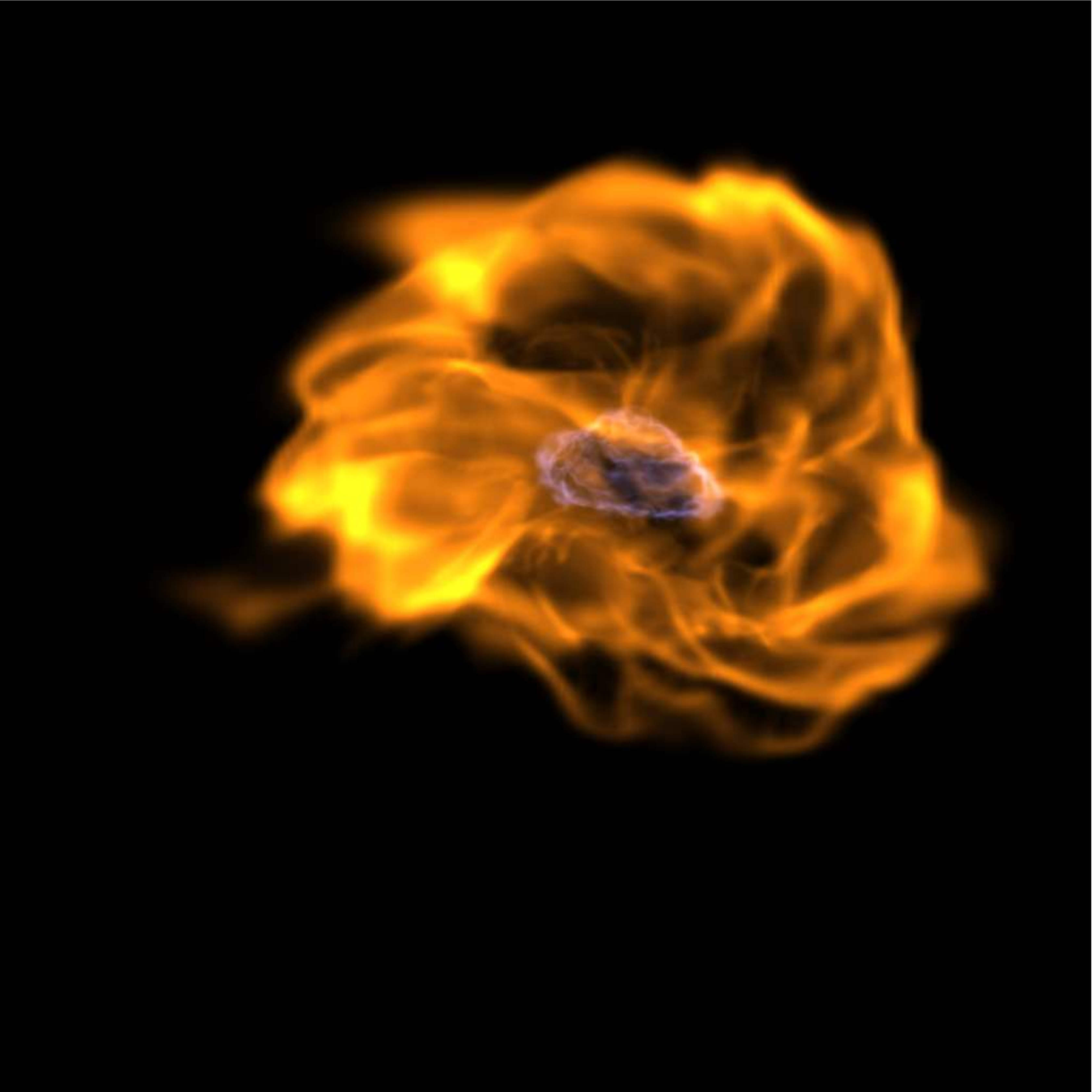}
\includegraphics[width=220pt]{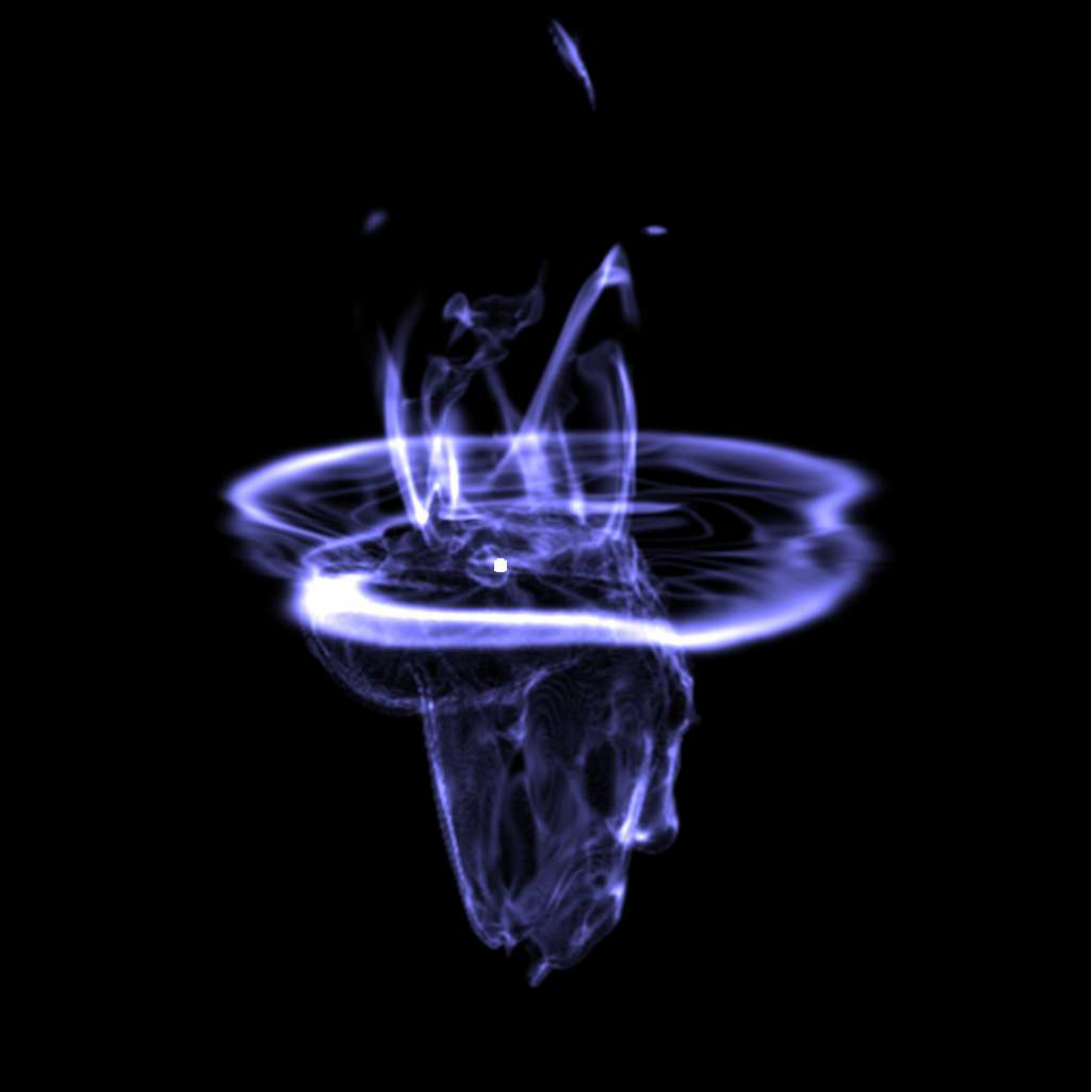}
\caption{\emph{(left)} Density isosurfaces of a primordial minihalo \citep{petersetal12c}. One can see the onion-like structure of the
gas density. \emph{(right)} Density isosurface of an ionisation-driven
molecular outflow \citep{petersetal10a}. The star that drives the outflow by its radiation feedback is shown in the centre of the image.}
\label{fig:simdata}
\end{figure}

Apart from its use to create analytical models, RADMC-3D can also be employed to post-process simulation data. We demonstrate this with two examples.
The left panel of Fig.~\ref{fig:simdata} shows density isosurfaces of a minihalo taken from a simulation of primordial star formation \citep{petersetal12c}.
Such minihalos typically collapse monolithically, creating an onion-like density structure very similar to the one in our analytical model.
We have used the same colours as in the previous section, but have adapted the normalisation constants for this particular situation. This time,
the higher densities are shown in purple and the lower ones in ochre. One can nicely see how the isosurfaces are deformed by turbulent motions
during the collapse of the halo.

The right panel of Fig.~\ref{fig:simdata} features a density isosurface of an ionisation-driven molecular outflow, taken from a collapse simulation of
massive star formation including ionisation feedback \citep{petersetal10a}. The isosurface shows both the disk in which the star forms as well
as the outflow driven perpendicular to it. A fourth frequency channel was used to display the star that drives the outflow, simply using a rectangle
function with a small radius around the star and drawing the resulting emission in the foreground of the image. One can see that the purple colour
becomes white were rays intersect the isosurface several times or where the isosurface is thicker than elsewhere. Again, this is because we are assuming a constant
thickness of the isosurface, which is clearly not the case. On the lower part of the image, some dark aliasing artifacts can be seen that are likely
due to our use of pre-classification.

Apart from the generation of nice images, this volume rendering module for RADMC-3D is also useful for diagnostic purposes because the transfer
functions can be made arbitrarily complex. For example, one might want to see an isosurface of density for all gas moving with a certain velocity,
coloured according to the temperature of the gas. Generating such a visualisation might be hard or even impossible with standard visualisation tools.

The integration of the radiative transfer equation for arbitrary transfer functions can also be used to measure physical quantities. For example,
using the opacities $\alpha_M = \rho$, $\alpha_P = \rho\, v$ and $\alpha_E = 0.5 \,\rho \,v^2$ with the gas density $\rho$ and velocity $v$ we can
compute the mass $M$, momentum $P$ and kinetic energy $E$ of the outflow by looking at the optical depth instead of the intensity.
These exact values can then be compared with figures derived using observational
techniques like synthetic CO line measurements \citep{petersetal12b}. This method is particularly useful when the outflows are misaligned
with the grid axes, so that a direct measurement from the simulation data is not straightforward.

\section{Conclusions}

I have pointed out an interesting connection between the physics of radiation transfer and the visualisation technique of direct volume rendering.
I have explained how the terminology used by computer scientists fits into the physical framework and described an implementation of direct
volume rendering in the astrophysical radiative transfer code RADMC-3D. I believe that the connection between computer graphics and radiation
transfer is an attractive topic for students and teachers not only of astrophysics, and that it can also help to better understand analytical models
and numerical simulations.

\section*{Acknowledgements}

It is a pleasure to thank Cornelis Dullemond for his ongoing support with the development of RADMC-3D
and for making this code freely available.
I acknowledge financial support through a Forschungskredit of the University of Z\"{u}rich, grant no. FK-13-112.

\def\newblock{\hskip .11em plus .33em minus .07em}

\end{document}